\documentstyle[aps,prl,twocolumn,epsfig,floats]{revtex}

\begin{document}

\wideabs{ 
\title{Tuning a Resonance in the Fock Space: Optimization of Phonon Emission in a
Resonant Tunneling Device}
\author{L. E. F. Foa Torres$^a$, H. M. Pastawski$^a$ \cite{Horacio} and S. S. Makler$%
^{b,c}$}
\address{$^a$Facultad de Matem\'{a}tica, Astronom\'{\i}a y F\'{\i}sica, Universidad\\
Nacional de C\'{o}rdoba, Ciudad Universitaria, 5000 C\'{o}rdoba, Argentina.}
\address{$^b$Instituto de F\'{\i}sica, Universidade do Estado do Rio de Janeiro,
21945-970 Rio de Janeiro, Brazil.}
\address{$^c$Instituto de F\'{\i}sica, Universidade Federal Fluminense, 
24210-340 Niter\'{o}i, Brazil.}
\date{\today}
\maketitle

\begin{abstract}
Phonon-assisted tunneling in a double barrier resonant tunneling device can
be seen as a resonance in the electron-phonon Fock space which is tuned by
the applied voltage. We show that the geometrical parameters can induce a
symmetry condition in this space that can strongly enhance the emission of
longitudinal optical phonons. For devices with thin emitter barriers this is
achieved by a wider collector barrier.
\end{abstract}



}


Progress in mesoscopic semiconductor devices \cite{cit-Marcus-Review} and
molecular electronics \cite{cit-Aviram-nature} is driven by the need of
miniaturization and the wealth of new physics provided by coherent quantum
phenomena. A fundamental idea behind these advances was Landauer's view that 
{\it conductance is transmittance }\cite{cit-Landauer,cit-Buettiker}.{\it \ }%
Hence, the typical conductance peaks and valleys, observed when some control
parameter is changed, are seen as fringes in an interferometer. However, the
many-body electron-electron (e-e) and electron-phonon (e-ph) interactions
restrict the use of this picture.  The e-e effects\ received much attention
in different contexts\cite{cit-Marcus-Review}. While interest on e-ph
interaction remained mainly focused on double barrier Resonant Tunneling
Devices (RTD) \cite{cit-Goldman-PRB}, where phonon-assisted tunneling shows
up as a satellite peak in a valley of the current-voltage (${\sf I}$-${\sf V}
$) curve,\  recent observation of electro-mechanical effects in molecular
electronics\cite{cit-Ho} requires a reconsideration of the e-ph problem.
Theory evolved from a many-body Green's function in a simplified model for
the polaronic states \cite{cit-Wingreen} to \ quantum and classical rate
equations approach \cite{cit-DATTA-e-ph}. The latter uses an incoherent
description of the e-ph interaction by adopting an imaginary self-energy
correction to the electronic states\cite{cit-Datta-Keldysh,cit-DAmato}.

In this article, we analyze a quantum coherent solution of transport with
e-ph interaction. We resort to a mapping of  the many-body problem into a
one-body scattering system where each phonon mode adds a new dimension to
the electronic variable \cite{cit-BrazJPhys,cit-Bonca}. Transmission of
electrons between incoming and outgoing channels with different number of
phonons are then used in a Landauer's picture where the only incoherent
processes occur inside the electrodes. This allows to develop the concept of 
{\it resonance in the e-ph Fock space} and the identification of the control
parameters that optimize the coherent processes leading to a maximized
phonon emission. It also gives a clue as to how ``decoherence'' arise within
an exact many-body description. As an application,\ we consider a RTD phonon
emitter where the relevant parameters are best known. There, the first
polaronic excitation\ serves as an ``intermediate'' state for the phonon
emission. An electron with kinetic energy $\varepsilon \leq \varepsilon _{F}$
and potential energy $e{\sf V}$ in the emitter {\it decays through tunneling
into} an electron with energy $\varepsilon +e{\sf V-}\hbar \omega _{0}$ in
the collector {\it plus} a longitudinal optical (LO) phonon. The tuning
parameter is the applied voltage while the optimization of phonon emission
requires the tailoring of the tunneling rates.\ 

\begin{figure}[tbp]
\centering \leavevmode
\center{\epsfig{file=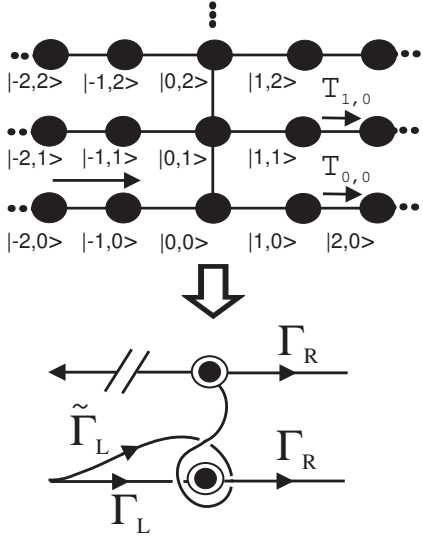, ,width=5.5cm,angle=0}}
\vspace{0.2cm}
\caption{Simple model: dots are states in the Fock space, lines are
interactions. The effective Hamiltonian including two entangled polaronic states is represented at the bottom.}
\label{fig-1}
\end{figure}

Let us consider a minimal Hamiltonian:

\begin{eqnarray}
{\cal H} &=&\sum_{j}%
\{E_{j}c_{j}^{+}c_{j}^{{}}-V_{j,j+1}(c_{j}^{+}c_{j+1}^{{}}+c_{j+1}^{+}c_{j}^{{}})\}+
\nonumber \\
&&+\hbar \omega _{0}b^{+}b^{{}}-V_{g}\,\,\,c_{0}^{+}c_{0}^{{}}(b^{+}+b^{{}}),
\label{eq-H}
\end{eqnarray}
where $c_{j}^{+}$and $c_{j}$ are electron creation and annihilation
operators at site $j$ on a 1-d chain with lattice constant $a$ and hopping
parameters $V_{j,j+1}=V.$ Tunneling rates are fixed by $V_{0,1}=V_{{\rm R}}$
and $V_{-1,0}=V_{{\rm L}}$ ($V_{{\rm L(R)}}\ll V$). The site energies are $%
E_{j}=2V$ for $j<0$ and $2V-e{\sf V}$\ for $j>0.$ $E_{0}$ $=E_{(o)}-\alpha $ 
$e{\sf V}$ is the well's {\it ground state} (including the charging effect)
shifted by the electric field. For barrier widths $L_{{\rm L}}$ and $L_{{\rm %
R}}$ and well size $L_{{\rm W}}$ a linear approximation for the potential
profile gives $\alpha =(L_{{\rm L}}+L_{{\rm W}}/2)/(L_{{\rm L}}+L_{{\rm W}%
}+L_{{\rm R}})$. We consider a single LO-phonon mode and an interaction ($%
V_{g}$) limited to the well. $b^{+}$ and $b$ are the creation and
annihilation operators for phonons. We restrict the Fock space to that
expanded by $\left| j,n\right\rangle =c_{j}^{+}\left( b^{+}\right) ^{n}/%
\sqrt{n!}\left| 0\right\rangle ,$ which maps to the 2-dimensional one-body
problem shown in Fig. \ref{fig-1}. The number $n$ of phonons is the vertical
dimension \cite{cit-BrazJPhys,cit-Bonca}. The horizontal dangling chains can
be eliminated through a decimation procedure \cite{cit-DAmato,cit-Levstein}
leading to an effective Hamiltonian: 
\begin{eqnarray}
\widetilde{{\cal H}}_{e-ph} &=&\sum_{n\geq 0}\{[E_{0}+n\hbar \omega
_{0}+\Sigma _{n}(\varepsilon )]\left| 0,n\right\rangle \left\langle
0,n\right| -  \label{eq-Heff-0,nbasis} \\
&&-\sqrt{n+1}V_{g}\left( \left| 0,n+1\right\rangle \left\langle 0,n\right|
+\left| 0,n\right\rangle \left\langle 0,n+1\right| \right) \},  \nonumber
\end{eqnarray}
The electron hopping into the electrodes is taken into account by the $%
\varepsilon $-dependence of the retarded self-energy corrections $\Sigma
_{n}=\,^{{\rm L}}\Sigma _{n}+\,^{{\rm R}}\Sigma _{n}$ . Specifically, $^{%
{\rm L}}\Sigma _{n}=\left| \frac{V_{{\rm L}}}{V}\right| ^{2}\times \Sigma
\,(\varepsilon -n\hbar \omega _{0})$, $^{{\rm R}}\Sigma _{n}=\left| \frac{%
V_{{\rm R}}}{V}\right| ^{2}\times \Sigma \,(\varepsilon -n\hbar \omega _{0}+e%
{\sf V})$, with:

\begin{eqnarray}
\Sigma (\varepsilon ) &=&\Delta \left( \varepsilon \right) -{\rm i}\Gamma
\left( \varepsilon \right) ;\text{ }\Delta \left( \varepsilon \right) =%
{\textstyle{1 \over \pi }}%
\int \frac{\Gamma \left( \varepsilon ^{\prime }\right) }{\varepsilon
-\varepsilon ^{\prime }}d\varepsilon ^{\prime },  \nonumber \\
\Gamma \left( \varepsilon \right) &=&\sqrt{V^{2}-\left( \varepsilon
/2+V\right) ^{2}}\times \theta (\varepsilon )\times \theta (4V-\varepsilon ).
\label{eq-general-sigmas}
\end{eqnarray}
While the imaginary part $\Gamma =\hbar v_{\varepsilon }/a$ is proportional
to the group velocity $v_{\varepsilon }$ in the electrodes, the actual
escape rates $\Gamma _{{\rm L(R)}}/\hbar $ are barrier controlled. For width 
$L_{{\rm L}({\rm R})}$ and attenuation length $\xi $ , $\Gamma _{{\rm L(R)}%
}/\Gamma =\left| V_{{\rm L(R)}}/V\right| ^{2}$ $\simeq \exp [-L_{{\rm L}(%
{\rm R})}/\xi ]$.

The retarded Green function connecting states $i$ and $n$, 
\begin{equation}
G_{n,i}^{R}\left( \varepsilon \right) =\left\langle 0,n\right| (\varepsilon 
{\cal I}-\widetilde{{\cal H}}_{e-ph}\left( \varepsilon \right) )^{-1}\left|
0,i\right\rangle ,  \label{eq-G}
\end{equation}
has poles at the exact eigenenergies. If $\Sigma _{n}(\varepsilon )\equiv 0$%
, these are the polaronic energies $E_{0}-\frac{\left| V_{g}\right| ^{2}}{%
\hbar \omega _{0}}+n\hbar \omega _{0}$. The transmission coefficient $%
T_{n,i} $ from the $i$-th incoming channel at left electrode to the $n$-th
channel at right is \cite{cit-DAmato}:

\begin{equation}
T_{n,i}^{{\rm \ }}\left( \varepsilon \right) =2%
\mathop{\rm Im}%
[^{{\rm R}}\Sigma _{n}\left( \varepsilon \right) ]\,\,\,\left|
G_{n,i}^{R}\left( \varepsilon \right) \right| ^{2}\,\,2%
\mathop{\rm Im}%
[^{{\rm L}}\Sigma _{i}\left( \varepsilon \right) ].
\label{eq-general-transmittance}
\end{equation}
If the Fermi energy $\varepsilon _{F}\ll V,$ Eq.(\ref{eq-general-sigmas})
becomes:

\begin{equation}
\Sigma \left( \varepsilon \right) \approx -{\rm i\,}\Gamma (\varepsilon
=\varepsilon _{F})\times \theta (\varepsilon ),  \label{eq-broad-band}
\end{equation}
and the $\theta $-function may cancel some $T$'s.

\begin{figure}[tbp]
\centering \leavevmode
\center{\epsfig{file=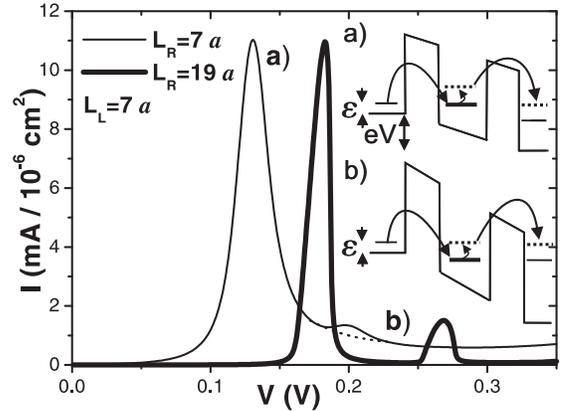, ,width=7.5cm,angle=0}}
\vspace{0.3cm}
\caption{Current density as a function of the applied voltage for a
symmetrical (thin line) and optimized (thick line) structures with $L_{{\rm L%
}}=7a$ ($19.7$ \AA ). The dotted line indicates the background current in
the region of the satellite peak for the symmetrical structure. The
inelastic processes contributing to the peaks a) and b) are represented in
the inset.}
\label{fig-2}
\end{figure}

To obtain the {\it elastic transmittance} when $g=$ $(V_{g}/\hbar \omega
_{0})^{2}\ll 1$ and $\left( \varepsilon _{F},\Gamma _{{\rm L}}+\Gamma _{{\rm %
R}}\right) <\hbar \omega _{0}$, we need: 
\begin{eqnarray}
G_{0,0}^{R} &\simeq &\frac{1-g}{\varepsilon -\overline{E}_{0}+{\rm i}[\Gamma
_{{\rm L}}+\Gamma _{{\rm R}}]}+  \nonumber \\
&&+\frac{g}{\varepsilon -[\overline{E}_{0}+\hbar \omega _{0}]+{\rm i}[%
\widetilde{\Gamma }_{{\rm L}}+\Gamma _{{\rm R}}]},  \label{eq-G00(e-ph)}
\end{eqnarray}
evaluated with the first two polaronic states. Here, $\widetilde{\Gamma }_{%
{\rm L}}=g\Gamma _{{\rm L}}$ and $\overline{E}_{0}=E_{0}-\frac{\left|
V_{g}\right| ^{2}}{\hbar \omega _{0}}.$ The first term contains the main
resonance associated to the build up of the polaronic ground state. The
second\ term contains a virtual exploration into the first polaronic
excitation. It is noteworthy that when $\Gamma =0,$ this Green function
would cancel out at an intermediate energy giving rise to an {\it %
antiresonance}\cite{cit-DA-P-W,cit-Levstein}. This \ concept{\it \ } extends
the spectroscopic Fano-resonances \cite{cit-Fano} to the problem of
conductance\cite{cit-Kaster-antiresonance}. \ For $g\ll 1,$ this effect is
less important and in the whole energy range, 
\begin{equation}
T_{0,0}^{{\rm \ }}\simeq \frac{4\Gamma _{{\rm L}}\Gamma _{{\rm R}}}{\left[
\varepsilon -\overline{E}_{0}\right] ^{2}+[\Gamma _{{\rm L}}+\Gamma _{{\rm R}%
}]^{2}}+{\cal O}(g)  \label{eq-T00(e-ph)}
\end{equation}
describes the main resonant elastic peak at $\varepsilon =\overline{E}_{0}$.

\ The {\it inelastic transmittance,}\ $T_{1,0}^{{\rm \ }}$, can be evaluated
from: 
\begin{eqnarray}
G_{1,0}^{R} &\simeq &\frac{-\frac{V_{g}}{\hbar \omega _{0}}}{\varepsilon -%
\overline{E}_{0}+{\rm i}[\Gamma _{{\rm L}}+\Gamma _{{\rm R}}]}+  \nonumber \\
&&+\frac{\frac{V_{g}}{\hbar \omega _{0}}}{\varepsilon -\left[ \overline{E}%
_{0}+\hbar \omega _{0}\right] +{\rm i}[\widetilde{\Gamma }_{{\rm L}}+\Gamma
_{{\rm R}}]}.  \label{eq-G10(e-ph)}
\end{eqnarray}
When $\varepsilon +{\rm e}{\sf V}>\hbar \omega _{0}$ escapes are enabled and
its poles involve the processes represented in the inset of Fig. \ref{fig-2}%
: a) The first term gives an inelastic transmittance at the main peak. b)
The second term provides a satellite peak at $\varepsilon =\overline{E}%
_{0}+\hbar \omega _{0}$, associated to a polaronic excitation followed by
its decay into an escaping electron and a phonon left behind. Around this
satellite peak:

\begin{equation}
T_{1,0}^{{\rm \ }}\simeq \frac{4\widetilde{\Gamma }_{{\rm L}}\Gamma _{{\rm R}%
}}{\left[ \varepsilon -\left( \overline{E}_{0}+\hbar \omega _{0}\right) %
\right] ^{2}+[\widetilde{\Gamma }_{{\rm L}}+\Gamma _{{\rm R}}]^{2}},
\label{eq-T10(e-ph)}
\end{equation}
showing that phonon emission is a {\it resonance in the Fock-space} (see
bottom of Fig. \ref{fig-1}). A maximal probability ($T_{1,0}^{{\rm \ }}=1$)
requires equal rates of formation and decay\cite{cit-DA-P-W}: $\widetilde{%
\Gamma }_{{\rm L}}=\Gamma _{{\rm R}},$ which in our RTD implies: 
\begin{equation}
L_{{\rm R}}\simeq L_{{\rm L}}+2\xi \ln [\frac{\hbar \omega _{o}}{V_{g}}].
\label{eq-optimal-length}
\end{equation}
Hence, thin barriers with this generalized symmetry condition have $T_{1,0}^{%
{\rm \ }}\simeq 1$ over a broad energy range.

The application of the Keldysh formalism\cite{cit-GLBE2} to our Fock-space
gives an electrical current ${\sf I}_{{\rm tot}}$ expressed as a balance
equation\cite{cit-Buettiker} in terms of the transmittances of Eq. (\ref
{eq-general-transmittance}) and the electrochemical potentials. The
experimental condition of high bias and low temperature ($e{\sf V}%
>\varepsilon _{F}\gg k_{B}{\sf T}$), \ rules out right-to-left flow, while $%
\hbar \omega _{o}>\varepsilon _{F}$ , enables the $\theta $ in Eq. (\ref
{eq-broad-band}) preventing inelastic reflection \ and overflow \cite
{cit-Kirczenow} of the final states. Thus, 
\begin{equation}
{\sf I}_{{\rm tot}}=\sum_{n}{\sf I}_{n};\,\,{\rm where\,}{\sf I}_{n}=(%
{\textstyle{2e \over h}}%
)\int_{0}^{\varepsilon _{F}}T_{n,0}^{{}}(\varepsilon ){\rm d}\varepsilon ,
\label{eq-total-current}
\end{equation}
\smallskip The ``decoherence'' introduced by the e-ph interaction on the
former single particle description can be now appreciated. One aspect, valid
even if $\hbar \omega _{o}\rightarrow 0,$ is that in Eq. (\ref
{eq-total-current}) the outgoing currents can not interfere. Another is \
the phase-shift fluctuations and ``broadening'' of the one-particle resonant
energy induced by the virtual processes in the elastic channel of Eq. (\ref
{eq-G00(e-ph)}).

At the satellite peak, the main elastic contribution to the current is
provided by the off-resonant tunneling through the ground state, i.e. ${\sf I%
}_{0}\simeq 
{\textstyle{2e \over h}}%
4\Gamma _{{\rm L}}\Gamma _{{\rm R}}\,\,\varepsilon _{F}/\left( \hbar \omega
_{0}\right) ^{2}$. The inelastic current determined by Eqs.(\ref
{eq-T10(e-ph)}) and (\ref{eq-total-current}) is 
\begin{eqnarray}
{\sf I}_{1} &\simeq &%
{\textstyle{e \over \hbar }}%
\frac{4\widetilde{\Gamma }_{{\rm L}}\Gamma _{{\rm R}}}{(\widetilde{\Gamma }_{%
{\rm L}}+\Gamma _{{\rm R}})}\times \left[ 
{\textstyle{2 \over \pi }}%
\arctan \left( 
{\displaystyle{\varepsilon _{F} \over 2(\widetilde{\Gamma }_{{\rm L}}+\Gamma _{{\rm R}})}}%
\right) \right]   \label{eq-simple-inelastic-current} \\
&\simeq &\left\{ 
\begin{array}{ccc}
{\textstyle{e \over \hbar }}%
4\widetilde{\Gamma }_{{\rm L}}\Gamma _{{\rm R}}/(\widetilde{\Gamma }_{{\rm L}%
}+\Gamma _{{\rm R}}) & {\rm for} & \varepsilon _{F}\gg (\widetilde{\Gamma }_{%
{\rm L}}+\Gamma _{{\rm R}}) \\ 
\frac{2e}{h}T_{1,0}\times \varepsilon _{F} & {\rm for} & \varepsilon _{F}\ll
(\widetilde{\Gamma }_{{\rm L}}+\Gamma _{{\rm R}})
\end{array}
\right. .{\rm \,}  \nonumber
\end{eqnarray}
The first line differs from the result of rate equations in \cite
{cit-DATTA-e-ph} by the factor in brackets, fundamental to resolve extreme
regimes. When $\varepsilon _{F}\gg (\widetilde{\Gamma }_{{\rm L}}+\Gamma _{%
{\rm R}})$ the inelastic current becomes geometry independent in the wide
range of $\varepsilon _{F}\gg \Gamma _{{\rm R}}>\widetilde{\Gamma }_{{\rm L}}
$. In the opposite case ${\sf I}_{1},$\ and\ hence the power emitted as
phonons $\hbar \omega _{0}{\sf I}_{1}/e,$ becomes determined by the
transmittance at resonance, which is maximized by the generalized symmetry
condition of Eq. (\ref{eq-optimal-length}).

Each current term, ${\sf I}_{n>0},$ contributes with $n$ useful phonons,
while ${\sf I}_{0}$'s energy degrades fully into electrode heating. Then,
one might seek a maximal ratio between the inelastic power ${\sf P}_{{\rm in}%
}$ and the total power ${\sf P}$, 
\begin{equation}
\eta =%
{\displaystyle{{\sf P}_{{\rm in}} \over {\sf P}}}%
=%
{\displaystyle{\hbar \omega _{0}\sum_{n>0}n%
{\displaystyle{{\sf I}_{n} \over e}} \over {\sf I}_{{\rm tot}}{\sf V}}}%
,  \label{eq-efficiency}
\end{equation}
which is the efficiency to transform the electric potential energy into
LO-phonon energy. At the voltage tuning the resonance at the satellite peak $%
{\sf V}_{0}\simeq (E_{(o)}-\left| V_{g}\right| ^{2}/\hbar \omega _{0}+\hbar
\omega _{0}-\varepsilon _{F}/2)/\alpha ,$ the lowest order of $\eta $ has
two factors ${\sf I}_{1}/({\sf I}_{0}+{\sf I}_{1})$ and $\hbar \omega _{0}/e%
{\sf V}_{0}$.\ The first is small for narrow barriers because non-resonant
tunneling dominates over phonon-assisted tunneling. For wide barriers, it
goes to one as $\widetilde{\Gamma }_{{\rm L}}+\Gamma _{{\rm R}}\rightarrow 0$%
.\ The second decreases with increasing right barrier's width because it
requires a higher ${\sf V}_{0}$. Thus, as $\Gamma _{{\rm R}}$ is decreased,
two effects compete: the switch from non-resonant to phonon assisted
resonant tunneling and an excess in the electronic kinetic energy in the
collector. Hence, as long as the left barrier is not extremely thin ($\Gamma
_{{\rm L}}>\hbar \omega _{0}$), $\eta $ can not depend much on geometry.
With this restriction in mind, a device designed for phonon production
should maximize the emitted power according to Eq. (\ref{eq-optimal-length}).

Let us compare these basic predictions with the numerical results of a
description involving geometry, voltage and energy dependences of a typical
RTD. A discrete 3-d model\ is defined in terms of the effective mass $%
m^{\ast }$ with $V$ $=\hbar ^{2}/(2m^{\ast }a^{2})$. The potential profile
for the diagonal energies $E_{j}$ is shown in the inset of Fig. \ref{fig-2}. 
$N_{{\rm L}}$, $N_{{\rm R}}$ and $N_{{\rm w}}$ are the number of sites in
the left and right barriers and the well, the associated widths are $%
L_{i}=N_{i}a$. For translational symmetry along the interface, we consider a
single phonon mode per transversal (parallel to the interface) state, with
frequency \thinspace $\omega _{0}$ and localized in the structure region.
While conservation of transverse electron's momentum might not be fully
realistic \cite{cit-Turley1993}, it constitutes a first approximation
yielding results consistent with the main experimental features. The current
components are obtained from (\ref{eq-total-current}) by integration over
the transversal modes. The parameters in our calculations are chosen to
simulate the case of a GaAs-AlGaAs structure. The effective mass $m^{\ast }$
is $0.067$ $m_{e}$, the LO phonon frequency $\hbar \omega _{0}=36$ meV, $a=$
2.825\ \AA , and the hopping parameter $V=7.125$ eV. A typical e-ph
interaction strength of $g\sim 0.1$ is obtained with $V_{g}\,\simeq 10$ meV.
For a well of 56.5\AA , barrier heights of $300$ meV and $\varepsilon _{F}$
= $10$ meV, the inclusion of $n\leq 3$ warrants good numerical convergence.

\begin{figure}[tbp]
\centering \leavevmode
\center{\epsfig{file=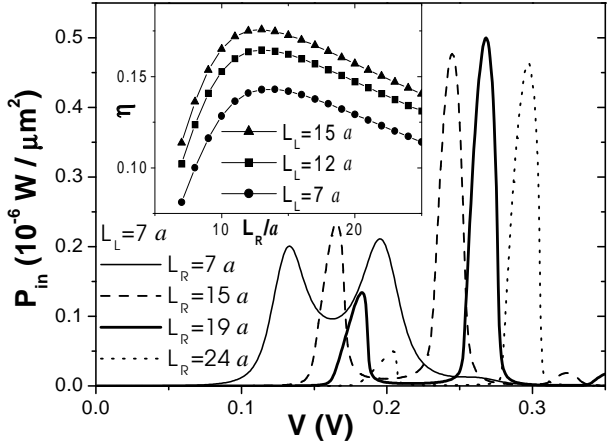, ,width=8.0cm,angle=0}}
\vspace{0.3cm}
\caption{Power emitted as LO phonons as a function of the applied voltage
for $L_{{\rm L}}=7a$\ ($19.7$ \AA ) and different values of $L_{{\rm R}}$.
The efficiency as a function of the right barrier width is shown in the
inset.}
\label{fig-3}
\end{figure}

For {\it wide left barriers} (of about 25$a\simeq $ 70\AA\ or more), we
found that the maximum value of ${\sf P}_{{\rm in}}$ varies slowly with the
width of the right barrier. Hence, consistently with our discussion of the
1-d model, there is no substantial gain in ${\sf P}_{{\rm in}}$ by choosing
an asymmetric structure. Consequently, a high phonon emission rate should be
sought for thin barriers.

A {\it thin left barrier} of $L_{{\rm L}}=7a$ (19.7\AA ), gives a tunneling
probability $T_{{\rm L}}(\varepsilon _{F})=\Gamma _{{\rm L}}/\Gamma \sim
0.03 $. Figure \ref{fig-2} \ shows the ${\sf I}${\sf -}${\sf V}$ curves for
symmetric and asymmetric RTDs. In Fig. \ref{fig-3}\ we show ${\sf P}_{{\rm in%
}}{\sf -V}$ for various right barrier widths $L_{{\rm R}}$. The peaks are
shifted to higher voltages as $L_{{\rm R}}$ is increased, because the
resonant energies are lowered approximately by $\alpha e{\sf V}$. We can
also see that the peak value of ${\sf P}_{{\rm in}}$ as a function of the
right barrier width exhibits a maximum. The ${\sf I}$-${\sf V}$ curve for
the optimal configuration is shown in Fig. \ref{fig-2} (heavy line). The
inset of Fig. \ref{fig-3} shows the dependence of $\eta ,$ evaluated at the
optimal voltage, on the right barrier width for various left barrier widths
where $\Gamma _{{\rm L}}<\hbar \omega _{0}$. In agreement with our
theoretical analysis, $\eta $ keeps the same magnitude for all the shown
geometrical configurations. The main result of Fig. \ref{fig-3} \ is the
confirmation that, for a given $L_{{\rm L}}$ satisfying $\Gamma _{{\rm L}%
}<\hbar \omega _{0}$ and $g\Gamma _{{\rm L}}+\Gamma _{{\rm R}}>\varepsilon
_{F}$, the phonon emission rate is enhanced by a factor 2.5 by choosing a
wider right barrier as prescribed by Eq.(\ref{eq-optimal-length}). This may
explain the unusually large satellite peaks of asymmetric structures\cite
{cit-Turley1993}.

An RTD optimized for phonon emission might have many applications. In fact,
in AlGaAs-GaAs RTD these primary LO\ phonons have a short life-time \cite
{cit-Vallee} and decay into a pair of LO and transverse acoustic (TA)
phonons. This phenomenon inspired the proposal \cite{cit-BrazJPhys} for the
generation of a coherent TA-phonon beam in an RTD (called a SASER)\cite
{cit-JPCM98}. That device {\it required} an energy difference between the
first two electronic states in the well $E_{1}-E_{0}=\hbar \omega _{0}$\cite
{cit-BrazJPhys,cit-JPCM98}. In contrast, the {\it present} proposal does not
require such an accurate device geometry. Instead, operation in the phonon
emission mode {\it only} requires the tuning of the many-body resonance with
the external voltage. Geometry just improves its yield by imposing a
generalized symmetry condition in the Fock-space. For a typical AlGaAs
emitter barrier of 20\AA\ this would require a 54\AA\ collector's barrier.
We expect that our results could stimulate the search for excited phonon
modes (e.g. with Raman spectroscopy), in operational RTD's as a function of
the applied voltage in the various configurations. While for simplicity we
have restricted our analysis to a model RTD, our analysis applies to other
problems \cite{cit-Ho} involving electronic resonant tunneling in the
presence of an interaction with an elementary excitation.

We acknowledge financial support from CONICET, SeCyT-UNC, ANPCyT and
Andes-Vitae-Antorchas. HMP and LEFFT are affiliated with CONICET.


\end{document}